\documentclass[doublecolumn]{ieeetran}

\usepackage{stfloats}
\usepackage{amsfonts}
\usepackage{amssymb}

\usepackage{amsthm}
\usepackage{cite}

\usepackage{float}
\usepackage{color}
\usepackage{stfloats,fancyhdr}
\usepackage{amsmath,bm}

\DeclareMathOperator*{\argmin}{arg\,min}

\usepackage{algorithm}
\usepackage{algorithmic}
\usepackage{multirow}
\usepackage{changepage}
\usepackage[normalem]{ulem}
\usepackage{amsthm}
\usepackage{balance}

\usepackage{bbm}

\newtheorem{theorem}{Theorem}
\newtheorem{lemma}{Lemma}

\newtheorem{definition}{Definition}
\newtheorem{assumption}{Assumption}
\newtheorem{remark}{Remark}

\usepackage{mathtools}
\mathtoolsset{showonlyrefs=true}

\usepackage{subfigure}
\usepackage{fancybox,dashbox}
\usepackage{authblk}
\usepackage{tabularx}

\usepackage{comment}

\newcommand{\myexpect}[1]{\mathsf{E}\left[#1\right]}

\newcommand{\myprob}[1]{\mathsf{Prob}\left[#1\right]}

\title{\LARGE \bf Stability Conditions for Remote State Estimation of Multiple Systems over Semi-Markov Fading Channels}
\author{
	Wanchun Liu, Daniel E.\ Quevedo,
	Branka Vucetic, Yonghui Li
}

\begin{document}
\maketitle

\thispagestyle{empty}
\pagestyle{empty}

\begin{abstract}
\let\thefootnote\relax\footnote{W.\ Liu, B.\ Vucetic and Y.\ Li are with School of Electrical and Information Engineering, The University of Sydney, Australia.
	Emails:	\{wanchun.liu,\ branka.vucetic,\ yonghui.li\}@sydney.edu.au. 
D. E.\ Quevedo is with the School of Electrical Engineering and Robotics, Queensland University of Technology (QUT), Brisbane, Australia.	Email: dquevedo@ieee.org.
}
This work studies remote state estimation of multiple linear time-invariant  systems over shared wireless time-varying communication channels. 
We model the  channel states by a semi-Markov process which captures both the random holding period of each channel state and the state transitions. The model is sufficiently general to be used in both fast and slow fading scenarios. 
We derive   necessary and sufficient stability conditions of the multi-sensor-multi-channel system in terms of the system parameters.
We further investigate how the delay of the channel state information availability and the holding period of channel states affect the stability. In particular, we show that, from a system stability perspective, fast fading channels may be preferable to slow fading ones.
\end{abstract}

\begin{IEEEkeywords}
Stability of linear systems, control over communications, estimation, Kalman filtering, Markov processes.
\end{IEEEkeywords}

\section{Introduction}
\IEEEPARstart{T}{he} incoming Fourth Industrial Revolution, Industry 4.0, 
focuses heavily on interconnectivity, automation, machine learning, and real-time data for customized and flexible mass production~\cite{Liu2021IoT}.
In particular, with low-cost and scalable deployment capabilities, 
wireless remote state estimation from ubiquitous sensors
will be essential in many industrial networked control applications, such as advanced manufacturing, warehouses automation, mining, and smart grids~\cite{ParkSurvey}.

The typical connection density in the Industry 4.0 era is about $10^6$/km$^2$; however, wireless communications 
have a limited spectrum bandwidth for transmissions.
Therefore, transmission scheduling among  sensors is a critical issue over the shared limited number of frequency channels.
Most wireless scheduling works focus solely on communications performance, including throughput, latency, and reliability, but are agnostic to upper-layer applications, such as estimation and control~\cite{sementic6G}.
However, for a multi-sensor-multi-channel remote estimation system, where each sensor measures an unstable dynamic plant, the scheduler must guarantee the stability of the remote estimation of all plant states. Otherwise, some of the plants cannot be stabilized, leading to catastrophic impacts on real-world systems.
The design of stabilizing multi-sensor-multi-channel remote estimators is a challenging problem and has drawn significant attention.

Optimal dynamic transmission scheduling policies of multi-plant networked systems over shared wireless resources were investigated in~\cite{gatsis2015opportunistic,Eisen}. 
However, the stability conditions of these systems have not been investigated.
Sensor transmission scheduling of remote estimation systems over single and multiple packet drop channels were investigated in~\cite{alexscheduling} and~\cite{Wu2018Auto}, respectively.
Once sufficient stability conditions  were obtained,   Markov decision process (MDP) methods were be adopted for finding the optimal scheduling policies.
The   work~\cite{LEONG2020108759} 
developed a sufficient stability condition over time-correlated fading channels, whereas~\cite{liu2021remote}   derived a necessary and sufficient stability condition.

Due to shadowing and multi-path propagation, wireless channel states (e.g., qualities) are time-varying and time-correlated~\cite{tse2005fundamentals}. Since wireless channel dynamics have significant impact on the remote estimation quality, accurate channel modeling is critical for stability analysis.
In the multi-system scheduling works above, time-invariant channels were considered in~\cite{alexscheduling,Wu2018Auto}; time-uncorrelated fading channels were adopted in~\cite{gatsis2015opportunistic,Eisen}; recently, more practical time-correlated fading channels were applied in \cite{LEONG2020108759,liu2021remote} and modeled by Markov processes.
However, the Markov modeling is suitable for fast fading channels, i.e., the channel state changes at each time, and is not accurate for  slow fading scenarios. 
As verified by experiments in industrial environments~\cite{semiMarkov}, semi-Markov processes, which  generalize  Markov processes, are suitable for characterizing slow fading channels in factories.
In addition, semi-Markov modeling also captures the time-varying feature of the channel state holding period in practice, i.e., the channel coherence time. Note that most of the existing channel models assume fixed coherence time for tractability~\cite{tse2005fundamentals}.

In this work, we focus on the stability   analysis of a multi-sensor remote estimation system over shared semi-Markov fading channels. The novel contributions include:

	$\bullet$ We build up a multi-sensor remote estimation system over practical multi-level semi-Markov fading channels.
	To the best of our knowledge, such a system has never been investigated in the open literature.
	Note that the existing works~\cite{LEONG2020108759,liu2021remote} only considered the simpler Markov channel modeling with binary-level channel states.
	
	$\bullet$  We derive  a necessary and sufficient stability condition for remote estimation and also provide the structure of a stability-guaranteeing scheduling policy. 
	Our result establishes a fundamental design guideline for stable remote estimation systems over practical wireless channels.
	
	$\bullet$  We also investigate how the delay of the channel state information availability and the holding period of channel states affects the stability. In particular, we show that a fast fading scenario (e.g., with a short average holding period) is preferable from a stability viewpoint.

\section{System Model}\label{sec:sys}
\begin{figure}[t]
	\centering\includegraphics[scale=0.45]{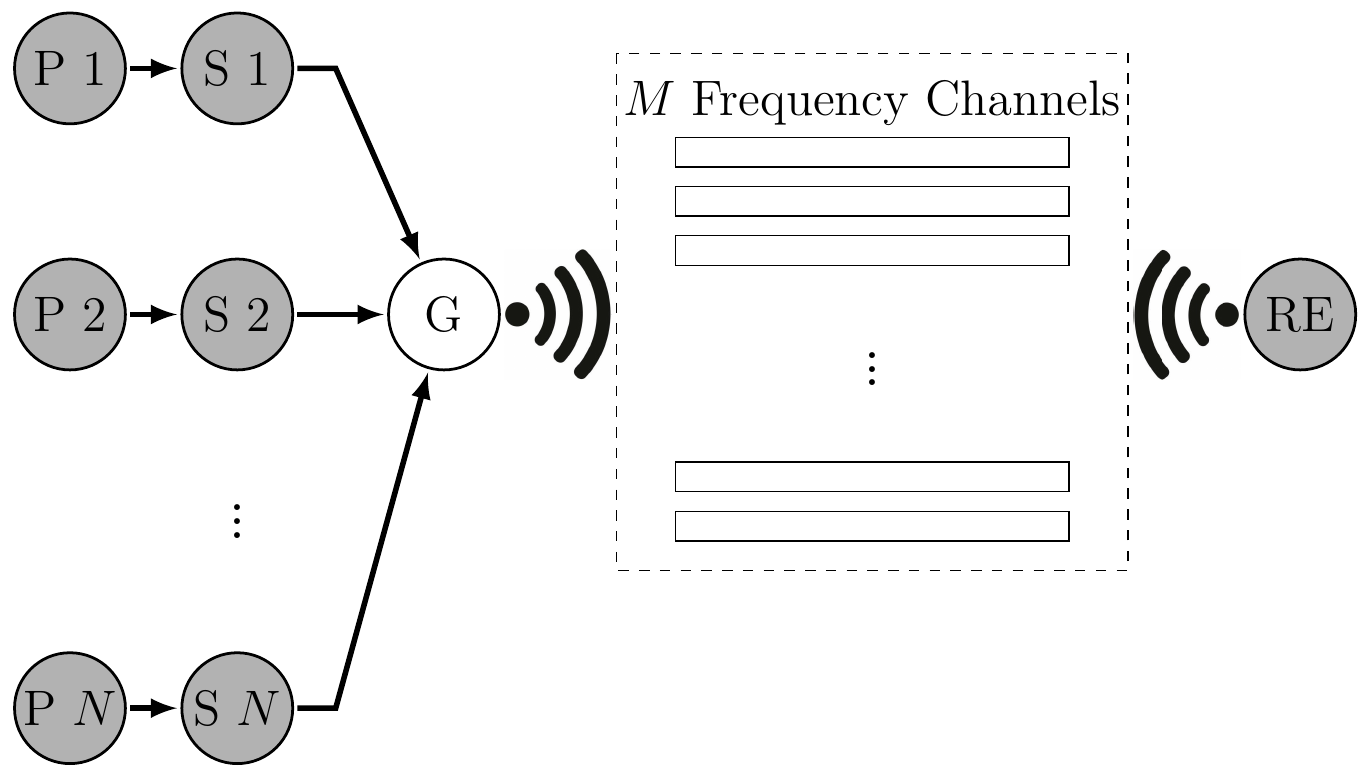}
	\caption{The multi-sensor-multi-frequency remote estimator. P, S, G, and RE denote processes, sensors, gateway and remote estimator, respectively.}
	\label{fig:system_model}
\end{figure}

We consider a remote estimation system with $N$ sensors each measuring an independent physical process, as illustrated in Fig.~\ref{fig:system_model}.
A local gateway connected to the sensors collects their  measurements and forwards them to a remote estimator.
Connections between sensors and the gateway are reliable and not scheduled, while the gateway to remote estimator communications are wireless and scheduled due to  bandwidth limitations.
There exist infinitely many dynamic transmission scheduling policies.
\emph{It is critical to determine necessary and sufficient conditions of the remote estimation system under which there exists a scheduling policy that can stabilize the system. }
If such a condition is not satisfied, then no stabilizing scheduling policy exists and one should redesign the system.
The main focus of the current letter is to present a necessary and sufficient stability condition, and thereby provide fundamental design guidelines for stable remote estimation systems.

Each process $n$ is modeled as an LTI system:
\begin{equation} \label{sys}
\begin{aligned}
\mathbf{x}_n{(t+1)} &= \mathbf{A}_n \mathbf{x}_n(t) + \mathbf{w}_n(t),\\
\mathbf{y}_n(t) &= \mathbf{C}_n\mathbf{x}_n(t) + {\mathbf{z}_n(t)},
\end{aligned}
\end{equation}
where 
$\mathbf{x}_n(t) \in \mathbb{R}^{l_n}$ and $\mathbf{y}_n(t) \in \mathbb{R}^{r_n}$ are the process state and the sensor measurements   at time $t\in\mathbb{N}_0$, respectively.
$\mathbf{A}_n \in \mathbb{R}^{{l_n} \times {l_n}}$ and $\mathbf{C}_n \in \mathbb{R}^{{r_n} \times {l_n}}$ are process $n$'s state transition matrix and  sensor $n$'s measurement matrix, respectively.
$\mathbf{w}_n(t) \in \mathbb{R}^{l_n}$ and $\mathbf{z}_n(t) \in \mathbb{R}^{r_n}$ are the process disturbance and the measurement noise, and are 
independent and identically distributed (i.i.d.) zero-mean Gaussian processes with the covariance matrices $\mathbf{W}_n$ and $\mathbf{Z}_n$, respectively.

\subsection{Local Estimation}\label{sec:local}
Each sensor uses a local Kalman filter (KF) to pre-process its measurement before sending to the gateway~\cite{LEONG2020108759}.
We have
\begin{equation}\label{sub:1}
	\begin{aligned}
	\mathbf{x}^s_n ({t|t-1})&=\mathbf{A}_n \mathbf{x}^s_n ({t-1})\\
	\mathbf{P}^s_n ({t|t-1})&=\mathbf{A}_n \mathbf{P}^s_n ({t-1}) \mathbf{A}^{\top}_n+\mathbf{W}_n\\
	\mathbf{K}_n (t)&=\!\mathbf{P}^s_n ({t|t-1}) \mathbf{C}^{\top}_n(\mathbf{C}_n \mathbf{P}^s_n ({t|t-1}) \mathbf{C}^{\top}_n\!+\!\mathbf{Z}_n)^{-1}\\
	\mathbf{x}^s_n (t)&=\!\mathbf{x}^s_n ({t|t-1})\!+\!\mathbf{K}_n (t)(\mathbf{y}_n({t})\!-\!\mathbf{C}_n \mathbf{x}^s_n ({t|t-1}))\\
	\mathbf{P}^s_n (t)&=(\mathbf{I}_n-\mathbf{K}_n ({t}) \mathbf{C}_n)\mathbf{P}^s_n ({t|t-1})
	\end{aligned}
\end{equation}
where $\mathbf{x}^s_n ({t|t-1})$ and $\mathbf{x}^s_n (t)$ are the predicted and updated state estimate at time $t$, respectively. $\mathbf{K}_n (t)$ is the Kalman gain. $\mathbf{P}^s_n ({t|t-1})$ and $\mathbf{P}^s_n (t)$ are the predicted and updated error covariance, respectively.
$\mathbf{I}_n$ is the $l_n \times l_n$ identity matrix
In particular, $\mathbf{x}^s_n (t)$ is the optimal estimate of $\mathbf{x}_n(t)$ at time $t$ in terms of the estimation mean-square error, where the estimation error covariance $\mathbf{P}^s_n (t)$ is defined as
\begin{equation}\label{eq:Ps}
\mathbf{P}^s_n(t) \triangleq \myexpect{(\mathbf{x}^s_n(t)-\mathbf{x}_n(t))(\mathbf{x}^s_n(t)-\mathbf{x}_n(t))^\top}.
\end{equation}
We assume that the \emph{local} KFs are stable and operate in steady state~\cite{LEONG2020108759,liu2020remote}, i.e., $\mathbf{P}^s_n(t) = \mathbf{\bar{P}}_n, \forall t\in \mathbb{N}_0, n \in \mathcal{N}\triangleq \{1,2,\dots,N\}$,
and investigate the \emph{remote} estimation stability.

\subsection{Semi-Markov Fading Channel}\label{sec:semi}
We assume that there are $M<N$ frequency channels for sensor data transmission. The channels are not perfect for transmission and can induce packet dropouts. The channel qualities at the $M$ frequencies  are time-correlated and modeled as a semi-Markov process as below. A better channel quality leads to a smaller packet drop probability.

Consider an $M$-frequency-multi-level channel quality state $\mathbf{h}(t) \triangleq [h_1(t),h_2(t),\dots,h_M(t)]$, where the quality of the $m$th frequency channel $h_m(t)$ has $M_m$ levels and 
$$h_m\!(t)\! \!\in\! \mathcal{H}_m \!\!\triangleq\! \{h^{(m)}_1,h^{(m)}_2,\dots,h^{(m)}_{M_m}\},\! \; m \!\in\! \mathcal{M} \!\triangleq\! \{1,\dots,M\}.$$ 
The packet drop probabilities of transmissions at different frequencies and different channel quality levels can be different.

\begin{figure}[t]
	\centering\includegraphics[scale=0.8]{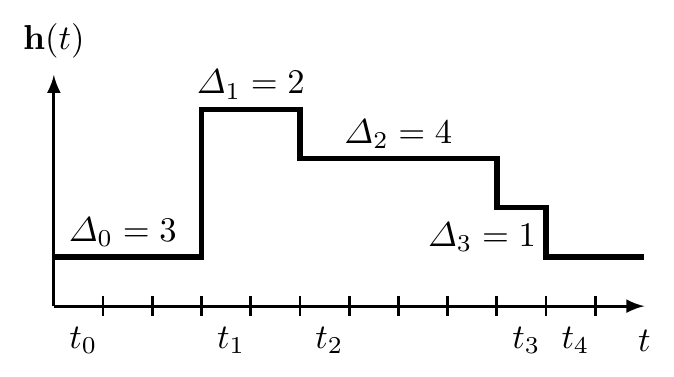}
	\caption{An illustration of the semi-Markov chain of $\{\mathbf{h}(t)\}$.}
	\label{fig:semi_markov}
\end{figure}

The channel quality state $\mathbf{h}(t)$ forms a semi-Markov chain with $\bar{M}=\prod_{m=1}^{M}M_m$ irreducible states, where
$$\mathbf{h}(t) \in \mathcal{S}  = \mathcal{H}_1 \times \mathcal{H}_2 \times \dots \times \mathcal{H}_M = \{\mathbf{h}_1,\mathbf{h}_2,\dots,\mathbf{h}_{\bar{M}}\}$$
and $\mathbf{h}_j = [h_{j,1},h_{j,2},\dots,h_{j,M}]$.
The transition instants between channel quality states are denoted by $\{t_0,t_1,\dots,t_l,\dots\}$, with $t_0=0$, and $t_0<t_1<t_2,\dots$ all integers.
The $l$th holding period, i.e., the amounts of time spent in the same channel quality state before the $l$th channel quality state transitions, is defined as $\varDelta_l \triangleq t_{l+1} -t_l$. Assuming that the holding periods are bounded by $\bar{\varDelta} \in \mathbb{N}$, we have $\varDelta_l \in 
\Delta \triangleq \{1,\dots,\bar{\varDelta}\}, \forall l$.
See Fig.~\ref{fig:semi_markov} for an illustration of the semi-Markov chain of $\{\mathbf{h}(t)\}$.
Note that the semi-Markov chain degrades to a Markov chain when $\bar{\varDelta} =1$.

Let $\mathbf{M} \in \mathbb{R}^{\bar{M}} \times \mathbb{R}^{\bar{M}}$ denote the channel quality state transition probability matrix of the semi-Markov chain, where the $i$th-row-$j$th-column element is
\begin{equation}\label{eq:Mij}
	[\mathbf{M}]_{i,j}\triangleq \myprob{\mathbf{h}(t_{l+1}) =\mathbf{h}_j \vert \mathbf{h}(t_l) =\mathbf{h}_i}, \forall l \in \mathbb{N}_0.
\end{equation}
The probability distribution of the holding period given the current channel quality state is
\begin{equation}\label{eq:holding_prob}
\psi_i(\delta) = \myprob{\varDelta_l= \delta \vert \mathbf{h}(t_l)=\mathbf{h}_i}, \forall l\in\mathbb{N}_0, \mathbf{h}_i\in\mathcal{S}.
\end{equation}
We assume that the channel quality transition and the holding period are independent, i.e.,
\begin{equation}
\myprob{\mathbf{h}(t_{l+1}) =\mathbf{h}_j,\varDelta_l = \delta \vert \mathbf{h}(t_l) =\mathbf{h}_i} =[\mathbf{M}]_{i,j} \psi_i(\delta).
\end{equation}

Let $\varDelta_l(t) \triangleq t - t_l+1$ denote the holding time of the current channel quality state, where $ t_l\leq t < t_{l+1}$.
Similarly, we define the holding time of the next channel quality state as $\varDelta_{l'}(t+1) \triangleq (t+1) - t_{l'}+1$, where $l\leq l' \leq l+1$.

From \eqref{eq:holding_prob}, the probability that the channel quality state transition occurs in the next time slot is
\begin{equation}\label{eq:stoping_con}
\begin{aligned}
&\myprob{\varDelta_{l'}(t+1)=1 \vert \varDelta_l(t)=\delta, \mathbf{h}(t) =\mathbf{h}_i} \\
&= \myprob{\varDelta_{l}=\delta \vert \varDelta_{l}\geq\delta, \mathbf{h}(t) =\mathbf{h}_i}
=\frac{\psi_i(\delta)}{\sum_{\delta'=\delta}^{\bar{\varDelta}} \psi_i(\delta')}.
\end{aligned}
\end{equation}

Now we define the channel state vector $\mathbf{\tilde{h}}(t)$ as the cascaded state of $\mathbf{h}(t)$ and
$\varDelta_l(t)$:
 $$
\mathbf{\tilde{h}}(t) \triangleq (\mathbf{h}(t),\varDelta_l(t))  \in \mathcal{\tilde{S}} \triangleq \mathcal{S} \times \Delta = \{\mathbf{\tilde{h}}_1,\mathbf{\tilde{h}}_2,\cdots,\mathbf{\tilde{h}}_{\tilde{M}}\},
$$ where the cardinality  $|\mathcal{\tilde{S}}| = \bar{M} \bar{\varDelta} \triangleq \tilde{M}$.
From \eqref{eq:Mij}, \eqref{eq:stoping_con}, and the semi-Markov property of $\{\mathbf{h}(t)\}$, it is easy to show that $\{\mathbf{\tilde{h}}(t)\}$ has the Markov property, i.e., given the current state $\mathbf{\tilde{h}}(t)$, the next state $\mathbf{\tilde{h}}(t+1)$ is independent of the previous  states $\{\mathbf{\tilde{h}}(0),\mathbf{\tilde{h}}(1),\dots,\mathbf{\tilde{h}}(t-1)\}$.
Thus, the original semi-Markov chain $\{\mathbf{h}(t)\}$ is converted to the Markov chain $\{\mathbf{\tilde{h}}(t)\}$.
In the rest of the paper, we will use $\mathbf{\tilde{h}}(t)$ in stead of $\mathbf{h}(t)$ for ease of analysis.

Using \eqref{eq:Mij}, \eqref{eq:holding_prob}, and \eqref{eq:stoping_con}, the channel state transition probability matrix $\mathbf{\tilde{M}} \in \mathbb{R}^{\tilde{M}} \times \mathbb{R}^{\tilde{M}}$ of $\{\mathbf{\tilde{h}}(t)\}$ can be obtained, and the $\tilde{i}$th-row-$\tilde{j}$th-column element is
\begin{equation}\label{eq:channel_transit}
\begin{aligned}
&[\mathbf{\tilde{M}}]_{\tilde{i},\tilde{j}} 
\!\triangleq\!
\myprob{\mathbf{\tilde{h}}(t+1)\!=\!\mathbf{\tilde{h}}_{\tilde{j}} \!=\! (\mathbf{h}_j,\delta_j) \vert \mathbf{\tilde{h}}(t) \!=\!\mathbf{\tilde{h}}_{\tilde{i}} \!=\! (\mathbf{h}_i,\delta_i)}\\
&=\begin{cases}
[\mathbf{M}]_{i,j} \frac{\psi_i(\delta_i)}{\sum_{\delta=\delta_i}^{\bar{\varDelta}} \psi_i(\delta)} &\text{ if } \delta_j=1,\\
1- \frac{\psi_i(\delta_i)}{\sum_{\delta=\delta_i}^{\bar{\varDelta}} \psi_i(\delta)} &\text{ if } \delta_j=\delta_i+1\leq \bar{\varDelta}, \mathbf{h}_j = \mathbf{h}_i, \\
0 &\text{ otherwise.}\\
\end{cases}
\end{aligned}
\end{equation}
For ease of analysis, we assume that $\{\mathbf{\tilde{h}}(t)\}$ is an aperiodic and irreducible Markov chain.

We make the channel state availability assumption as below.
\begin{assumption}[Known Current Channel State]\label{ass:cur-channel}
	At time $t\in \mathbb{N}$, the current channel state $\mathbf{\tilde{h}}(t)$ is known by the gateway prior to transmission scheduling.
\end{assumption}
\noindent We note that the channel state can be estimated based on standard channel estimation techniques~\cite{tse2005fundamentals}.
The scenario with delayed channel state information will be investigated at the end of Section~\ref{sec:key}.

We define the transmission failure event at frequency $m$ given the  the channel state  $\mathbf{\tilde{h}}_j$ as
$\mu\left(\mathbf{\tilde{h}}_j, m \right)=0$ and the packet drop probability
\begin{equation}\label{eq:error_rate}
d_{j,m} \triangleq \myprob{\mu\left(\mathbf{\tilde{h}}_j, m \right)=0} \in [0,1], \forall j\in\mathcal{\tilde{M}}, m\in\mathcal{M},
\end{equation}
where $\mathcal{\tilde{M}} \triangleq \{1,2,\dots,\tilde{M}\}$.

\subsection{Remote Estimation}\label{sec:remote}
In each time slot, the gateway collects $N$ packets of the sensor estimates $\{\mathbf{\hat{x}}^s_1(t),\dots,\mathbf{\hat{x}}^s_N(t)\}$, schedules $M$ of them, and sends through $M$ frequency channels to the remote estimator. 
Each frequency channel can transmit at most one packet at a time, and the unscheduled packets are discarded.
Each scheduled packet can take at most one frequency channel for transmission.

Due to the transmission scheduling and packet dropouts, the remote estimator cannot receive all sensor packets at each time. 
Let $\gamma_n(t)=1$ denote the event that sensor $n$'s packet is successfully received by the remote estimator at time $t$.
We also define the \emph{age-of-information} (AoI) for each sensor,
$\phi_n(t)\in \mathbb{N}$, which is the time duration between the previous successful sensor $n$'s packet detection and the current time $t$, i.e.,
$
\phi_n(t) \triangleq \min_{\{t': t'<t, \bm{1}(\gamma_n(t')=1),t'\in \mathbb{N}_0\}}  (t- t'),
$
where $\bm{1}(\cdot)$ is the indicator function.
Thus, the AoI state has the updating rule below
\begin{equation} \label{eq:AoI_update}
\phi_n(t) = \begin{cases}
1 &\text{ if } \gamma_n(t-1) =1\\
\phi_n(t-1)+1 &\text{ otherwise.} \\
\end{cases}
\end{equation}

The optimal \emph{minimum mean-square error} (MMSE) remote estimator~\cite{liu2020remote} works as below, considering the one-step transmission delay~\cite{BATTILOTTI2019108499}
\begin{equation}\label{eq:remote}
\hat{\mathbf{x}}_n(t) = \begin{cases}
\mathbf{A}_n {\mathbf{x}}_n^s(t-1),& \text{ if } \gamma_n(t-1)=1,\\
\mathbf{A}_n \hat{\mathbf{x}}_n(t-1),& \text{ otherwise.}
\end{cases}
\end{equation}
and can be simplified as
\begin{equation}\label{general_estimater}
\hat{\mathbf{x}}_n(t) = \mathbf{A}^{\phi_n(t)}_n {\mathbf{x}}^s_n(t-\phi_n(t)).
\end{equation}
From \eqref{eq:Ps} and \eqref{general_estimater}, the estimation error covariance of process $n$ is derived~as
\begin{align} 
\mathbf{P}_n(t) 
&\triangleq  \mathsf{E}\left[(\hat{\mathbf{x}}_n(t)-\mathbf{x}_n(t))(\hat{\mathbf{x}}_n(t)-\mathbf{x}_n(t))^{\top}\right]
= \theta^{\phi_n(t)}_n(\bar{\mathbf{P}}_n), \label{general_form}
\end{align}
where $\bar{\mathbf{P}}_n$ was defined in Section~\ref{sec:local}, 
$\theta_n(\mathbf{X}) = \theta_n^{1}(\mathbf{X}) \triangleq \mathbf{A}_n\mathbf{X}\mathbf{A}_n^{\top}+\mathbf{W}_n
$, and
$\theta_n^{m+1}(\cdot)  \triangleq \theta_n (\theta_n^{m}(\cdot)), m \in \mathbb{N}. $
Thus, the remote estimation quality of process $n$ at time~$t$ can be quantified via the sum average estimation error $\mathsf{E}\left((\hat{\mathbf{x}}_n(t)-\mathbf{x}_n(t))^\top (\hat{\mathbf{x}}_n(t)-\mathbf{x}_n(t))\right) =\text{Tr}\left(\mathbf{P}_n(t) \right)$, where $\text{Tr}(\cdot)$ is the trace operator. 
By introducing the following function
\begin{equation}\label{eq:c}
c_n(i)\triangleq \text{Tr}\left(\theta^i_n(\bar{\mathbf{P}}_n)\right), \forall i\in \mathbb{N}
\end{equation}
and using \eqref{general_form}, we have
\begin{equation} \label{trace}
\text{Tr}\left(\mathbf{P}_n(t) \right) \triangleq c_n(\phi_n(t)),
\end{equation}
which is the \emph{estimation cost function} of process $n$ and is determined by its AoI state $\phi_n(t) \in \mathbb{N}$.

Note that due to the transmission scheduling and error, the AoI state $\phi_n(t)$  can have unbounded support, i.e., the remote estimator may not receive sensor $n$'s packet for an arbitrarily long time. Thus, the cost function $c_n(\phi_n(t))$ in \eqref{eq:c} takes values from a countably infinite set $\{\text{Tr}\left(\theta^1_n(\bar{\mathbf{P}}_n)\right),\text{Tr}\left(\theta^2_n(\bar{\mathbf{P}}_n)\right),\dots\}.$

As discussed in ~\cite{liu2020remote}, if $\rho(\mathbf{A}_n)\geq 1$, the cost grows up unbounded with the increasing AoI. Our focus is on the remote estimator's stochastic stability defined as below.

\begin{definition}[Average Mean-Square Stability]
	The $N$-sensor-$M$-frequency remote estimation system described above is average mean-square stable, if the long-term average estimation cost $J$ is bounded, where
	\begin{equation}\label{longterm}
	J\triangleq \sum_{n=1}^{N} J_n,\ 
	J_n \triangleq \limsup_{T\to\infty}\frac{1}{T}\sum_{t=1}^{T} c_n(\phi_n(t)),n\in\mathcal{N}.
	\end{equation}
	
\end{definition}

Using~\cite[Lemma 1]{liu2021remote}, it is straightforward to establish the following property of $c_n(\cdot)$:
\begin{lemma}\label{lem:c}
	For any given $\epsilon>0$, there exists positive constants $\kappa$ and $\eta$ such that 
\begin{align}\label{eq:upper}
	&\text{\normalfont (Upper bound) } c_n(i) < \kappa \left(\rho^2(\mathbf{A}_n)+\epsilon \right)^i,\\ \label{eq:lower}
	&\text{\normalfont (Lower bound) } c_n(i) \geq \eta (\rho(\mathbf{A}_n))^{2i}, \forall i\in\mathbb{N}.
\end{align}
\end{lemma}

\subsection{Transmission Scheduling Policy}
We solely focus on deterministic stationary scheduling policies.
Let $\bm\nu(t) \triangleq [\nu_1(t), \dots,\nu_N(t)] \in \{0,1,\dots,M\}^N$ denote the scheduling action for the $N$ sensors at time $t$. In particular, if $\nu_n(t) =0$, sensor $n$ is not scheduled; if $\nu_n(t) \in \mathcal{M}$, sensor $n$ is scheduled at frequency $\nu_n(t)$.
The scheduling actions are sent to the sensors via feedback channels. We assume that these transmissions are error-free due to the small communication overhead.

From \eqref{trace}, the AoI state $\bm \phi(t) \triangleq [\phi_1(t),\phi_2(t),\dots,\phi_N(t)]$ determines the current estimation cost. From \eqref{eq:channel_transit} and \eqref{eq:remote}, the current channel state $\mathbf{\tilde{h}}(t)$ reflects the chances of transmission success in the current time slot, and will affect the next channel state and hence the estimation cost in the next step.
Then, using the Markov properties \eqref{eq:channel_transit} and \eqref{eq:AoI_update}, a scheduling policy $\pi(\cdot)$ should take into account both the channel and AoI states for decision making, i.e., 
\begin{equation}\label{eq:schedule_0}
\bm\nu(t) = \pi(\bm\phi(t),\mathbf{\tilde{h}}(t)).
\end{equation}


\section{Stability Conditions} \label{sec:key}
The LTI system model and the semi-Markov channel statistics jointly determine the stability of the overall remote estimator. 
Our result is stated in terms of the channel state transition matrix $\mathbf{\tilde{M}}$,
the length-$\tilde{M}$ channel selection vector
$\mathbf{v} \triangleq [v_1,v_2,\dots,v_{\tilde{M}}] \in \mathcal{M}^{\tilde{M}}$, where the $i$th element $v_i \in\mathcal{M}$ denotes the selected frequency index given the channel state $\mathbf{\tilde{h}}_i$, and the $\tilde{M}\times \tilde{M}$ diagonal packet drop probability matrix $\mathbf{V}(\mathbf{v})$ given the channel selection vector $\mathbf{v}\in \mathcal{M}^{\tilde{M}}$: 
\begin{equation}\label{eq:V}
\begin{aligned}
[\mathbf{{V}}(\mathbf{v})]_{m,m} 
&\triangleq d_{m,v_m},m \in \mathcal{\tilde{M}},
\end{aligned}
\end{equation}
where $d_{i,j}$ was defined in~\eqref{eq:error_rate}.

\begin{theorem}\label{theo:main}
	The $M$-sensor-$N$-frequency remote estimator described in Section~\ref{sec:sys} under Assumption~\ref{ass:cur-channel} can be stabilized if and only if the following condition holds
	\begin{equation}\label{eq:key1}
	\rho^2_{\max} \lambda<1,
	\end{equation}
	where $\rho_{\max} \triangleq \max_{n\in \mathcal{N}} \rho(\mathbf{A}_n)$ is the largest spectral radius of all processes, 
	\begin{equation}\label{eq:lambda_inf1}
	\lambda \triangleq \rho\left(\mathbf{V}(\mathbf{v}^\star) \mathbf{\tilde{M}}\right),\ \mathbf{v}^\star = [v^\star_1,\dots,v^\star_{\tilde{M}}],
	\end{equation}
	and 
\begin{equation} \label{eq:theo_v}
v^\star_i = \argmin_{m\in\mathcal{M}} d_{i,m}, i\in\mathcal{\tilde{M}}.
\end{equation}
\end{theorem}

\begin{remark}
We see that the stability depends on the system parameter of the most unstable process, the channel state dynamics, and the packet drop probabilities at different channel states.
Although Theorem~\ref{theo:main} does not provide direct insights on the structure of a stable scheduling policy,
we will construct a policy with stability guarantees in the proof of the sufficient condition. Numerical examples of the stability condition are provided in Section~\ref{sec:num}.
\end{remark}

We will prove the necessary and sufficiency parts of Theorem~\ref{theo:main} in the sequel. Note that if all processes are stable, i.e., $\rho_{\max}<1$, the remote estimator is always stable. Thus, in the following, we only focus on the case with $\rho_{\max}\geq 1$.

\begin{proof}[Proof of Necessity]
	The proof has three parts: 1) the  construction of a virtual policy that can always achieve an average cost of the remote estimator lower than any real scheduling policy; 2) the average cost function analysis of the virtual policy;  3) the derivation of the necessary condition.
\subsubsection{Policy construction}\label{sec:nec_policy}
To prove the necessity,
we consider a virtual scenario that only the packet of the sensor corresponding to the most unstable process is scheduled for transmission in each time slot, while the other sensors' estimates are perfectly known by the remote estimator and need not packet transmissions. 
Without loss of generality, we assume that process~$1$ is the most unstable one, i.e., $\rho_{\max}=\rho(\mathbf{A}_1)$.
In other words, only sensor $1$'s is scheduled in each time slot and it can select any of the frequencies for transmission.  
For ease of notation, we will drop out the process index $n$ in the following analysis.
Furthermore, we replace the cost function $c(i)$ with its lower bound~\eqref{eq:lower}.

Thus, the original sensor scheduling policy~\eqref{eq:schedule_0} is reduced to a frequency selection one
\begin{equation}\label{eq:schedule_1}
\nu(t) = \pi(\phi(t),\mathbf{\tilde{h}}(t)) \in \mathcal{M},
\end{equation}
where $\nu(t)$ and $\phi(t)$ are the frequency selection action and the AoI state of sensor $1$, respectively. 
Recall that we focus on deterministic stationary scheduling policies, and hence drop the time index $t$ in the following.

Let $v_i(\phi) \triangleq \pi(\phi,\mathbf{\tilde{h}}_{i})\in\mathcal{M}$ denote the selected channel for transmission given the current AoI state $\phi$ and the channel state $\mathbf{\tilde{h}}_{i}$.
From \eqref{eq:schedule_1}, for given $\phi$, the frequency selection rule at $\tilde{M}$ different channel states can be uniformly written as:
\begin{equation}
\mathbf{v}(\phi) = [v_1(\phi),\dots, v_{\tilde{M}}(\phi)]\in \mathcal{M}^{\tilde{M}}.
\end{equation}
Given  $\phi$ and the frequency selection vector $\mathbf{v}(\phi)$, we obtain the packet drop probability matrix $\mathbf{V}(\mathbf{v}(\phi))$ based on \eqref{eq:V}:
\begin{equation}\label{eq:Ev2}
\begin{aligned}
[\mathbf{V}(\mathbf{v}(\phi))]_{m,m} = d_{m,v_m(\phi)}, \forall m \in \mathcal{\tilde{M}}.
\end{aligned}
\end{equation}

From \eqref{eq:AoI_update}, the current frequency selection action $\nu$ will only affect the next AoI state $\phi'$ and cost $c(\phi')$, and has no impact on the next channel state.
Given the current AoI $\phi$, the current channel state $\mathbf{\tilde{h}}_i$, and the selected frequency $m$,
the probabilities that $\phi'=\phi+1$ and $1$ are $d_{i,m}$ and $1-d_{i,m}$, respectively.
Using the monotonicity of the cost function in \eqref{eq:lower}, the AoI state $1$ is a better state than $\phi+1$ in terms of the current and future cost functions.
So the frequency selection action that leads to the lowest chance of state $\phi+1$ is the best action in terms of the long-term average cost.
The optimal frequency selection policy is a greedy one that always select the frequency with the lowest packet drop probability at each time, and thus is independent of the AoI state.

\begin{lemma} \label{lem:greedy}
For the remote estimator described above, if there exists deterministic and stationary frequency selection policies that stabilize the remote estimator, the optimal policy achieving the minimum average cost is independent of the AoI state, and is given by
	\begin{equation}\label{eq:freq_selec}
	\nu^\star_i = \pi^\star(\mathbf{\tilde{h}}_i) \triangleq \argmin_{m\in\mathcal{M}} d_{i,m}, \forall i \in \mathcal{\tilde{M}}.
	\end{equation}
\end{lemma}
In what follows, we solely need to analyze the average cost of the remote estimator over the optimal policy above. 

\subsubsection{Analysis of the average cost and the necessary condition}\label{sec:nec_analysis}
We adopt an estimation cycle based analysis  method that we developed earlier in~\cite{liu2020remote,liu2021remote}.
The infinite time horizon are divided into estimation cycles, each starting after a successful transmission and ending at the next one. Thus, the AoI state is always equal to $1$ at the beginning of each estimation cycle, and linearly increases step-by-step.
Let $T_k$ denote the length of the $k$th estimation cycle.
$C_k$ is the sum cost in the $k$th estimation cycle and is a function of $T_k$ as
\begin{equation}\label{g_fun}
C_k = g(T_k) \triangleq \sum_{j=1}^{T_k} \eta (\rho(\mathbf{A}_n))^{2j} \geq  \eta (\rho(\mathbf{A}_n))^{2T_k}.
\end{equation}

Define the channel state before the $k$th cycle as $\mathbf{b}_k\in \mathcal{\tilde{S}}$.
Without loss of generality,  assume that the first $\tilde{M}_1$ channel states of $\mathcal{\tilde{S}}$ can be pre-cycle states, where $0 < \tilde{M}_1 \leq \tilde{M}$, i.e., not all channel states in $\mathcal{\tilde{S}}$ have to be a pre-cycle state.
The channel state $\mathbf{\tilde{h}}_i$ with a strictly zero chance of transmission success can never be a pre-cycle state, i.e., $\min_{m\in\mathcal{M}} d_{i,m}=1$.
Similar to  \cite[Lemma~1]{liu2021remote}, the Markovian property of the pre-cycle channel states below can be proved.
\begin{lemma}\label{lem:G}
	$\{\mathbf{b}\}_{\mathbb{N}}$ is a time-homogeneous ergodic Markov chain with $\tilde{M}_1\leq \tilde{M}$ irreducible states of $\mathcal{\tilde{S}}$. 
	The state transition matrix of $\{\mathbf{b}\}_{\mathbb{N}}$ is $\mathbf{G}'$, which is the $\tilde{M}_1$-by-$\tilde{M}_1$ matrix taken from the top-left corner of
\begin{equation}
\mathbf{G} = \sum_{j=1}^{\infty} \mathbf{\tilde{\Xi}}(j),
\end{equation}
where
$
\mathbf{\tilde{\Xi}}(j)= \mathbf{\mathbf{\Xi}}(j-1)\left(\mathbf{I} - \mathbf{V}^\star\right)\mathbf{\tilde{M}}, j=1,2,\dots,
$
$\mathbf{V}^\star \triangleq \mathbf{V}(\mathbf{v}^\star)$, and 
\begin{equation}\label{eq:Xi}
\mathbf{\mathbf{\Xi}}(j)= 
\begin{cases}
\mathbf{I},& j =0\\
 (\mathbf{V}^\star \mathbf{\tilde{M}})^j,& j >0.
\end{cases}
\end{equation}
\end{lemma}
We note that the terms $\mathbf{\Xi}(j-1)$ and  $\left(\mathbf{I} - \mathbf{V}^\star\right)\mathbf{\tilde{M}}$ in $\mathbf{\tilde{\Xi}}(j)$ are related to the $(j-1)$ times of consecutive failed transmissions and the successful transmission right after these of a length-$j$ estimation cycle.
Let $\boldsymbol{\beta} \triangleq [\beta_1,\dots,\beta_{\tilde{M}_1}]^\top$ denote the stationary distribution of the Markov chain $\{\mathbf{b}\}_{\mathbb{N}_0}$, which is the unique null-space vector of $(\mathbf{I-G'})^\top$ and $\beta_i >0,\forall i\in \mathcal{\tilde{M}}_1$, where $\mathcal{\tilde{M}}_1\triangleq \{1,2,\dots, \tilde{M}_1\}$.

Due to the ergodicity of $\{\mathbf{b}_k\}$ and the definition in \eqref{g_fun}, it directly follows that the random processes $\{T_k\}$ and $\{C_k\}$ are ergodic. Using the property that the time average is equal to the ensemble average of an ergodic process, we drop the time index of $T_k$, $C_k$ and $\mathbf{b}_k$, and have
$
\myexpect{C} =\lim\limits_{K\rightarrow \infty }\frac{1}{K} \sum_{k=1}^{K} C_k =  \sum_{m=1}^{\tilde{M}_1} \beta_m \myexpect{C \vert \mathbf{b}=\mathbf{\tilde{h}}_m}.
$

By following the same steps in the proof of Theorem 1 of~\cite{liu2021remote}, we can first show that the average cost $J < \infty$ if and only if $\myexpect{C} < \infty$, and then derive the necessary condition making  $\myexpect{C}$ bounded as $\rho^{2}(\mathbf{A}) \rho\left(\mathbf{V}^\star\mathbf{\tilde{M}}\right)<1$.
\end{proof}

\begin{proof}[Proof of Sufficiency]
	We construct a \emph{persistent serial scheduling} policy that persistently schedules sensor 1's transmission at a time until it is successful and then schedules sensor $2$ and so on. For each transmission, the frequency is selected based on the scheme~\eqref{eq:freq_selec}.
Using the upper bound of the per-step cost function~\eqref{eq:upper} and following the similar analytical steps of the average sum cost per estimation cycle in~\cite{liu2021remote}, we can derive the upper bound of each sensor's average sum cost per cycle and then obtain the sufficient condition of Theorem~\ref{theo:main}.
\end{proof}

\begin{remark}
Although the policy constructed above   is a stability-guaranteeing one, it does not utilize the parallel frequency channels and is strictly not optimal.
For the optimal policy design, once the stability condition is satisfied, we can first design a suitable MDP problem, and then use classic dynamic programming or deep reinforcement learning algorithms to solve it, see for example~\cite{LEONG2020108759,liu2021DRL}.
\end{remark}

\textbf{{Extension Scenario}}:
We also investigate the stability condition of the scenario with delayed channel state information.
\begin{assumption}[Known Previous Channel State~\cite{LEONG2020108759,liu2021remote}]\label{ass:pre-channel}
	At time $t$, only the previous channel state $\mathbf{\tilde{h}}(t-1)$ is available.
\end{assumption}

Building on the multi-level Markov channel modeling of $\{\mathbf{\tilde{h}}(t)\}$ in Section~\ref{sec:semi} and following the same analytical steps in~\cite{liu2021remote}, which focused on the binary-level Markov channel scenario under Assumption~\ref{ass:pre-channel}, we can derive the stability condition as below.
\begin{theorem}\label{theo:main1}
	The $M$-sensor-$N$-frequency remote estimator described in Section~\ref{sec:sys} under Assumption~\ref{ass:pre-channel} can be stabilized if and only if
	$
	\rho^2_{\max} \lambda_\infty <1,
	$
	where
$
\lambda_{\infty} \triangleq \min_{L\in\mathbb{N}} \lambda_L,
$
\begin{equation}\label{eq:lambda_main}
\lambda_L \triangleq \min_{\mathbf{v}_l \in \mathcal{M}^{\tilde{M}}} \rho\left(\mathbf{E}(\mathbf{v}_1) \mathbf{E}(\mathbf{v}_2) \cdots \mathbf{E}(\mathbf{v}_L)\right)^{\frac{1}{L}}
\end{equation}
and $\mathbf{E}(\mathbf{v})$ is an $\tilde{M}\times \tilde{M}$ matrix generated by the channel selection vector $\mathbf{v} = [v_1,\dots,v_{\tilde{M}}]\in \mathcal{M}^{\tilde{M}}$ as 
\begin{equation}
\begin{aligned}
&[\mathbf{E}(\mathbf{v})]_{i,j} \triangleq 	\myprob{\mathbf{\tilde{h}}(t) = \mathbf{\tilde{h}}_j, \mu\left(\mathbf{\tilde{h}}_j, v_i \right) =0 \vert \mathbf{\tilde{h}}(t-1) = \mathbf{\tilde{h}}_i}\\
&=\mathbf{\tilde{M}}_{i,j} d_{j,{v_i}}.
\end{aligned}
\end{equation}
\end{theorem}

%
\begin{remark}\label{coro:compare}
	The stability condition in Theorem~\ref{theo:main1} is more restrictive than that in Theorem~\ref{theo:main}.
	Intuitively, the  gap in stability is introduced by the delay of the channel state information for scheduling.
	The scheduler with current channel state information can make better decision for stabilizing the system than the one with outdated information.	
	
	We will show that $\lambda$ in \eqref{eq:lambda_inf1} is no larger than $\min_{L\in\mathbb{N}} \lambda_L$ in \eqref{eq:lambda_main}.		
	Using the property that $\rho(\mathbf{X}^L)^{1/L} = \rho(\mathbf{X})$, for any square matrix $\mathbf{X}$ and positive integer $L$, we have
	$
	\lambda = \rho\left(\left(\mathbf{V}(\mathbf{v}^\star) \mathbf{\tilde{M}}\right)^L\right)^{1/L} , \forall L.
	$
	From the definition of $\mathbf{V}(\mathbf{v}^\star)$ in \eqref{eq:theo_v}, the following inequality about non-negative matrices holds element-wise
	$
	\mathbf{V}(\mathbf{v}^\star) \mathbf{\tilde{M}}
	\preccurlyeq 
	\mathbf{E}(\mathbf{v}), \forall \mathbf{v} \in \mathcal{M}^{\tilde{M}},
	$
	and hence
	$
	\left(\mathbf{V}(\mathbf{v}^\star)\mathbf{\tilde{M}}\right)^L \preccurlyeq \mathbf{E}(\mathbf{v}_1)\mathbf{E}(\mathbf{v}_2)\cdots\mathbf{E}(\mathbf{v}_L), \forall \mathbf{v}_1,\dots,\mathbf{v}_L.
	$
	Using the property that the spectral radius of one non-negative matrix is larger than the other if the former is larger   element-wise, it is readily   shown that $\lambda\leq \lambda_L, \forall L$, and hence $\lambda\leq \lambda_\infty$.
\end{remark}

\section{Numerical Examples and Conclusions}\label{sec:num}
In practice, the maximum frequency number of WiFi can be $48$, and the maximum holding period can be $100$ calculated by the practical channel coherence time and packet duration. For the illustration of the stability condition, we consider a simple 3-sensor-2-frequency remote estimation system with a maximum holding period $\bar{\varDelta}=2$.

The spectral radii of the three processes are $\rho(\mathbf{A}_1)=1.5$, $\rho(\mathbf{A}_2)=1.2$, and $\rho(\mathbf{A}_3)=1.1$.
Each of the two frequency channels has two quality states,   $\mathcal{H}_1 = \{h^{(1)}_1,h^{(1)}_2\}$ and $\mathcal{H}_2 = \{h^{(2)}_1,h^{(2)}_2\}$.
The packet drop probability at frequency-$m$-state-$i$ is denoted as $\tilde{d}_{m,i}$. 
Thus, the (vector) channel quality state has $\bar{M}=4$ states in total, $\mathbf{h}(t)\in\{\mathbf{h}_1=[h^{(1)}_1,h^{(2)}_1],\mathbf{h}_2=[h^{(1)}_1,h^{(2)}_2],\mathbf{h}_3=[h^{(1)}_2,h^{(2)}_1],\mathbf{h}_4=[h^{(1)}_2,h^{(2)}_2]\}$.
The channel quality transition probability  is $\mathbf{M}=[0.1\ 0.2\ 0.3\ 0.4; 0.2\ 0.1\ 0.4\ 0.3; 0.4\ 0.2\ 0.1\ 0.3; 0.3\ 0.1\ 0.4\ 0.3]$.
The probability distribution of the holding period is 
\begin{equation}
\psi_i(\delta)= 
\begin{cases}
\tilde{\psi}_1 &\text{ if } \delta=1\\
\tilde{\psi}_2 = 1-\tilde{\psi}_1 &\text{ if } \delta=2
\end{cases},
\forall i=1,2,3,4. 
\end{equation}
Thus, the  state $\mathbf{\tilde{h}}(t)=(\mathbf{h}(t),\varDelta_l(t))$ has $ \tilde{M} =\bar{M} \times \bar{\varDelta} =8$ states:  $(\mathbf{h}_1,1), (\mathbf{h}_1,2),\dots,(\mathbf{h}_4,1), (\mathbf{h}_4,2)$.
From~Theorem~\ref{theo:main}, the $8$ diagonal elements of the matrix $\mathbf{V}(\mathbf{v}^\star)$ are $
\min\{\tilde{d}_{1,1},\tilde{d}_{2,1}\}$, $\min\{\tilde{d}_{1,1},\tilde{d}_{2,1}\}$, 
$\min\{\tilde{d}_{1,1},\tilde{d}_{2,2}\}$, 
$\min\{\tilde{d}_{1,1},\tilde{d}_{2,2}\}$, 
$\min\{\tilde{d}_{1,2},\tilde{d}_{2,1}\}$, 
$\min\{\tilde{d}_{1,2},\tilde{d}_{2,1}\}$, 
$\min\{\tilde{d}_{1,2},\tilde{d}_{2,2}\}$, and  $\min\{\tilde{d}_{1,2},\tilde{d}_{2,2}\}$.
From \eqref{eq:channel_transit}, the channel state transition matrix $\mathbf{\tilde{M}}$ can be obtained directly.

\begin{figure}[t]
	\centering
	\includegraphics[scale=0.68]{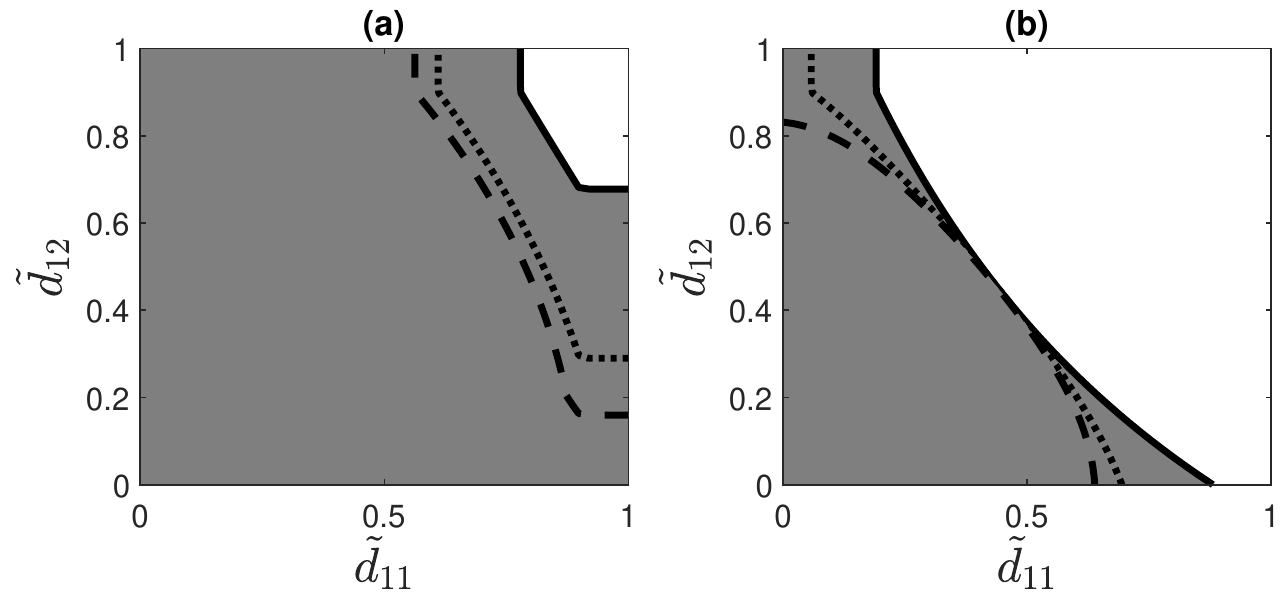}
	\caption{Stability regions (gray colored) with $\tilde{d}_{21} = 0.2, \tilde{d}_{22} = 0.9$ in (a) and $\tilde{d}_{21} = 0.9, \tilde{d}_{22} = 0.9$ in (b), where the solid, doted, and dashed lines indicate the regions with $\tilde{\psi}_1=0.99, 0.5$, and $0.1$, respectively.}
	\label{fig:region1}
\end{figure}

In Fig.~\ref{fig:region1}(a), we plot the stability regions in terms of $\tilde{d}_{11}$ and $\tilde{d}_{12}$ based on Theorem~\ref{theo:main}, where $\tilde{d}_{21}=0.2$ and $\tilde{d}_{22}=0.9$.
It is interesting to see that the stability region increases with $\tilde{\psi}_1$, i.e., the probability that the holding period of the channel condition is $1$.
This implies that a fast fading scenario can lead to a better stability than a slow fading one. 
Compared to Fig.~\ref{fig:region1}(a), we increase the packet drop probability at frequency-$2$-state-$1$, i.e., $\tilde{d}_{2,1}$, from $0.2$ to $0.9$ in Fig.~\ref{fig:region1}(b). We see that the reduced transmission reliability has lead to diminished stability regions as expected.

\section{Conclusions}\label{sec:con}
We have investigated the  necessary and sufficient stability conditions of the multi-plant remote estimation system.
For future work, in addition to the local filter-based remote estimation scenario, we will also consider the extension to transmission of raw measurements and investigate the stability conditions.
%

\balance
    


\end{document}